\documentclass{mn2e}
\usepackage{psfig}

\def\ltsima{$\; \buildrel < \over \sim \;$}
\def\lsim{\lower.5ex\hbox{\ltsima}}
\def\gtsima{$\; \buildrel > \over \sim \;$}
\def\gsim{\lower.5ex\hbox{\gtsima}}
\def\mes{M\'esz\'aros}

\def\mos{Mochkovitch}
\def\bsax{{\it Beppo}SAX}

\begin{document}

\title[BATSE GRB precursors]{Precursor activity in bright long BATSE 
$\gamma$-Ray Bursts}

\author[Davide Lazzati]
{Davide Lazzati$^{1,2}$\\ 
$^1$ JILA, Campus Box 440, University of Colorado, Boulder, CO 80309-0440 \\
$^2$ Institute of Astronomy, University of Cambridge,
Madingley Road, Cambridge CB3 0HA, England \\ 
{\tt e-mail: lazzati@quixote.colorado.edu} }

\maketitle

\begin{abstract}
We study a sample of bright long BATSE GRB light curves in the $200$~s
before the detection of the GRB prompt emission. We find that in a
sizable fraction of cases ($\sim 20\%$) there is evidence of emission
above the background coming from the same direction of the GRB. This
emission is characterised by a softer spectrum with respect to the
main one and contains a small fraction ($0.1-1\%$) of the total event
counts. The precursors have typical delays of several tens of seconds
extending (in few cases) up to 200 seconds (the limit of the
investigated period). Their spectra are typically non-thermal
power-law but for a few cases. Such long delays and the non-thermal
origin of their spectra are hard to reconcile with any model for the
progenitor.
\end{abstract}

\begin{keywords}
gamma-ray: bursts --- radiation mechanisms: non thermal
\end{keywords}

\section{Introduction}

In most Gamma-Ray Burst (GRB)models the main event is anticipated by a
less intense emission, characterised by a thermal spectrum, called a
precursor. Precursors can be distinguished into fireball precursors
and progenitor precursors. The former are associated with the moment
in which the fireball undergoes a transition from optical thickness to
optical thinness (Paczynski 1986; Lyutikov \& Usov 2000; \mes\ \& Rees
2000; Daigne \& \mos\ 2002; Lyutikov \& Blandford 2004), while the
latter are associated to the interaction of the jet with the
progenitor star (especially in the context of massive star
progenitors: Ramirez-Ruiz, MacFadyen \& Lazzati 2002; Waxman \&
Meszaros 2003). The discovery of precursors, the understanding of
their origin and the study of their properties would be of great
importance to constrain the physics and some parameters of the burst
outflow.

In the fireball precursor scenario, measuring the delay, duration,
luminosity and typical frequency of the precursor would allow one to
solve its properties and derive the Lorentz factor of the ejecta,
their temperature at the transparency radius and the transparency and
internal shock radii. All this parameters are extremely hard to
measure otherwise. On the other hand, should a precursor be recognised
as a progenitor precursor, its properties would give important
constraints on the dynamics of the jet propagation in the progenitor
and on the size of the progenitor star.

Observationally, little has been done so far. The main problem in
identifying a precursor to a GRB lies in the definition itself of what
a precursor is. As a matter of fact, defining a precursor implies
defining a starting point of the main emission different from ``the
first photon I see''. This problem has been addressed differently in
the past by different authors. Koshut et al. (1995) defined precursors
all the GRB pulses that had a peak intensity lower than that of the
whole burst and were followed by a period of quiescience longer than
the remaining burst active time. They searched the BATSE lightcurves
for precursor activity and found that $\sim3\%$ of all the BATSE GRB
lightcurves showed signs of precursor activity. Their precursors were
bright, containing a sizable fraction of the total event counts.

Since the precursors are theoretically predicted to be soft, another
approach can be used with instruments sensitive in the keV range. In
this case, it is possible to define a precursor as an emission episode
that is present in the low-energy instrument but not in the high
energy ones that are used to define the trigger of the main
event. Soft precursors were detected with GINGA (GRB~900126: Murakami
et al. 1991), HETE2 (GRB030329; Vanderspek et al. 2004) and \bsax\
(GRB011121; Piro et al. in preparation). The GINGA precursor is the
only one with a thermal spectrum, so that a comparison with
theoretical expectations could be performed (Ramirez-Ruiz et
al. 2002). The \bsax\ and HETE2 precursors are both inconsistent with
thermal emission and fitted successfully with a power-law.

In this paper, we perform a search for precursor activity in bright
long BATSE GRB lightcurves, with a different approach with respect to
Koshut et al. (1995). First, as a first condition to be considered
precursor activity, an emission event must be detected \emph{before}
the GRB trigger. This is a somewhat instrumental definition, but since
we are looking for weak precursors (theory predicts that precursors
should contain a very small fraction of the burst energy), it turns
out to be an effective one. Second, we ask that the episode we call a
precursor should decay in flux before the trigger. This choice is set
in order to exclude slowly rising GRB emission from the precursor
sample. We find, after rejecting some spurious events as background
fluctuations, that many more GRBs are characterised by this kind of
precursor activity ($\sim20\%$ with respect to the $\sim3\%$ of Koshut
et al. 1995). We also find a-posteriori that our precursor definition
is a good one, since these events have different properties than the
main emission.  Most of our precursors are characterised by
non-thermal spectra and long delays, both properties being hard to
reconcile with any model for their production.

This paper is organised as follows: in \S~2 we describe the selection
of our sample of GRB lightcurves; in \S~3 we detail the data analysis
and checks performed to build the precursor catalogue; in \S~4 we
describe the properties of the precursors and their relation to the
properties of the main GRB and in \S~5 we discuss our results.

\section{Sample selection}

The search was performed on a sub-sample of the BATSE GRB catalogue
tailored to contain bright and energetic events, belonging to the
sub-class of long-soft bursts. We selected therefore from the final
BATSE GRB catalogue\footnote{{\tt
http://www.batse.msfc.nasa.gov/batse/grb/catalog/current/} see also
Paciesas et al. (1999).} all the events with duration $T_{90}\ge5$~s,
Fluence ${\cal{F}}\ge1.62\times10^{-5}$~erg~cm$^{-2}$ and $1.024$~s
integrated peak flux $F_{\rm{pk}}\ge3.2$~cts~cm$^{-2}$~s$^{-1}$. We
also excluded bursts that overlapped with either weaker of stronger
bursts. This led to a sample of 146 bursts, 13 of which had to be
rejected because of a non-continuous coverage of the light curve in
the interval $t_{\rm{GRB}}-250<t<t_{\rm{GRB}}+T_{90}$, where
$t_{\rm{GRB}}$ is the trigger time of the considered GRB. This led to
a final sample of 133 GRB light curves.

\section{Data analysis}

The search for precursor activity was performed in several
steps. First, for each burst in the sample, the {\tt DISCLA}
continuous data file was retrieved from the archive\footnote{\tt
ftp://cossc.gsfc.nasa.gov/compton/data/batse/daily}. From the file,
data for the 8 detector and the 4 channels were independently
extracted for the interval
$t_{\rm{GRB}}-300<t<t_{\rm{GRB}}+1.5\,T_{90}$, amounting to 32 light
curves for each burst.

\begin{figure*}
\psfig{file=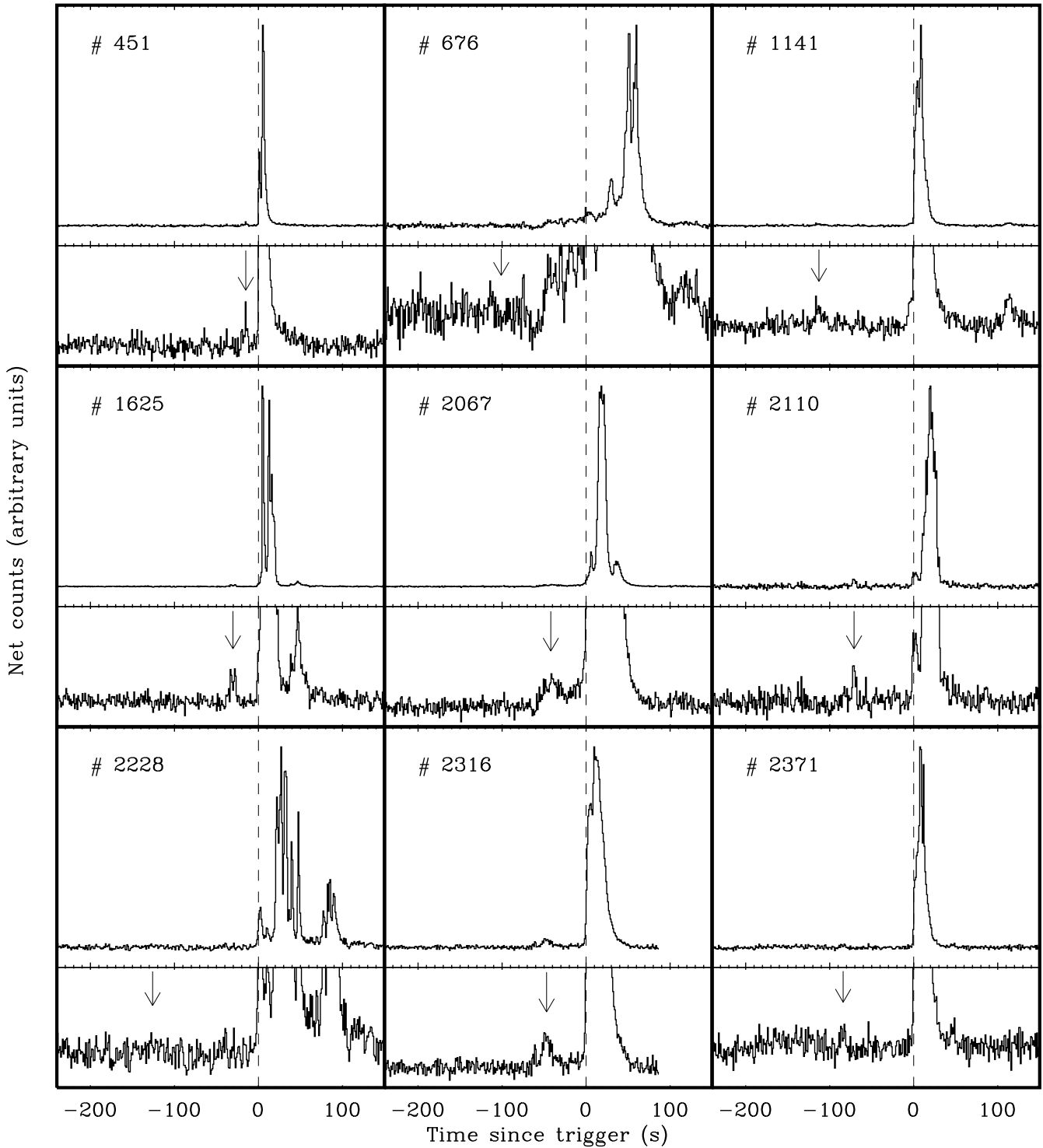,width=\textwidth}
\setcounter{figure}{0}
\caption{{Atlas of the GRB lightcurves for which a precursor activity 
has been detected. For each GRB, in the upper panel the full light
curve is shown, while in the lower panel the y-axis is zoomed in order
to emphasise the precursor emission, indicated with arrow(s). The
vertical dashed line shows the trigger time.}
\label{fig:atlas}}

\end{figure*}
\begin{figure*}
\psfig{file=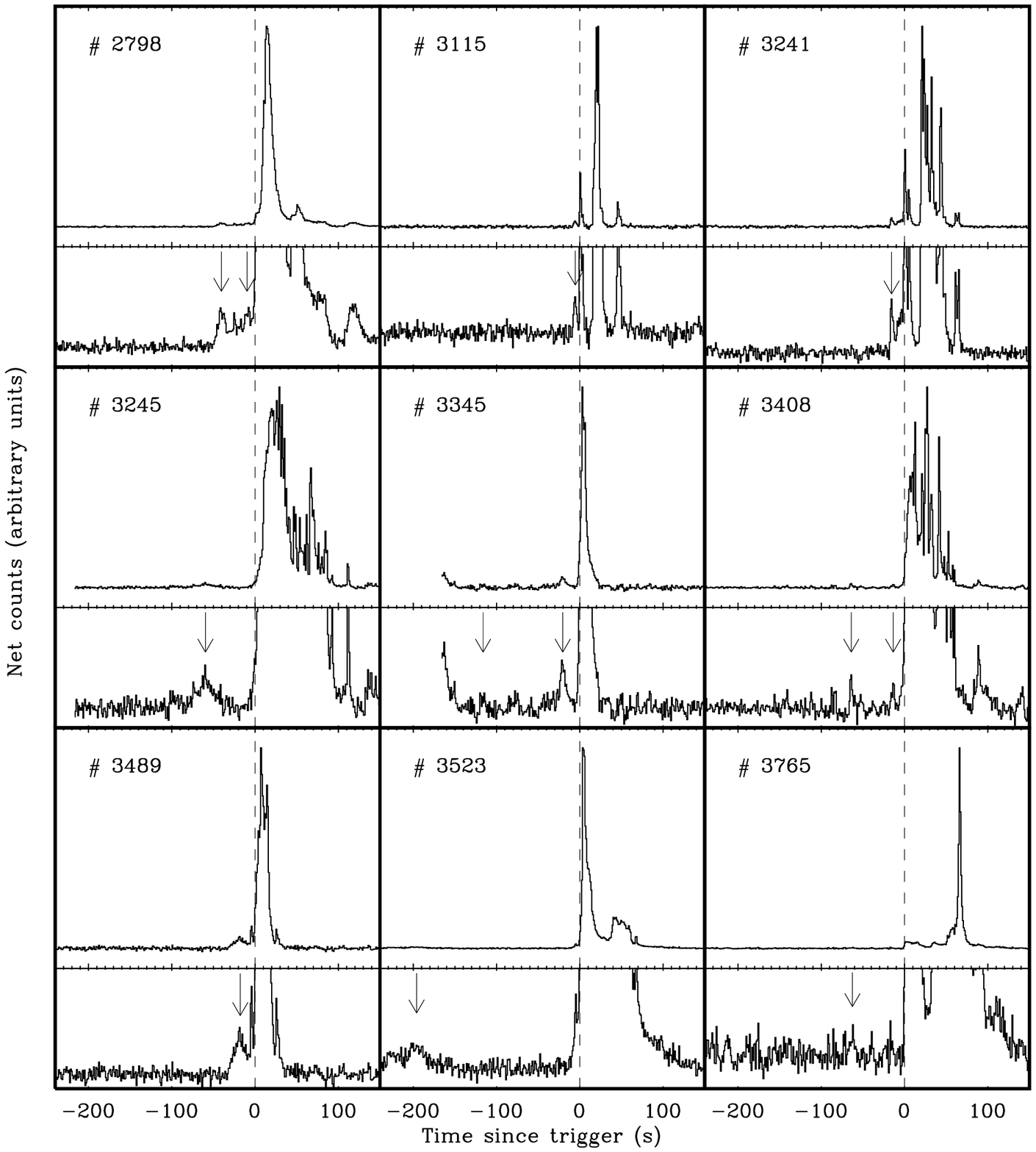,width=\textwidth}
\setcounter{figure}{0}
\caption{{continued}}
\end{figure*}

\begin{figure*}
\psfig{file=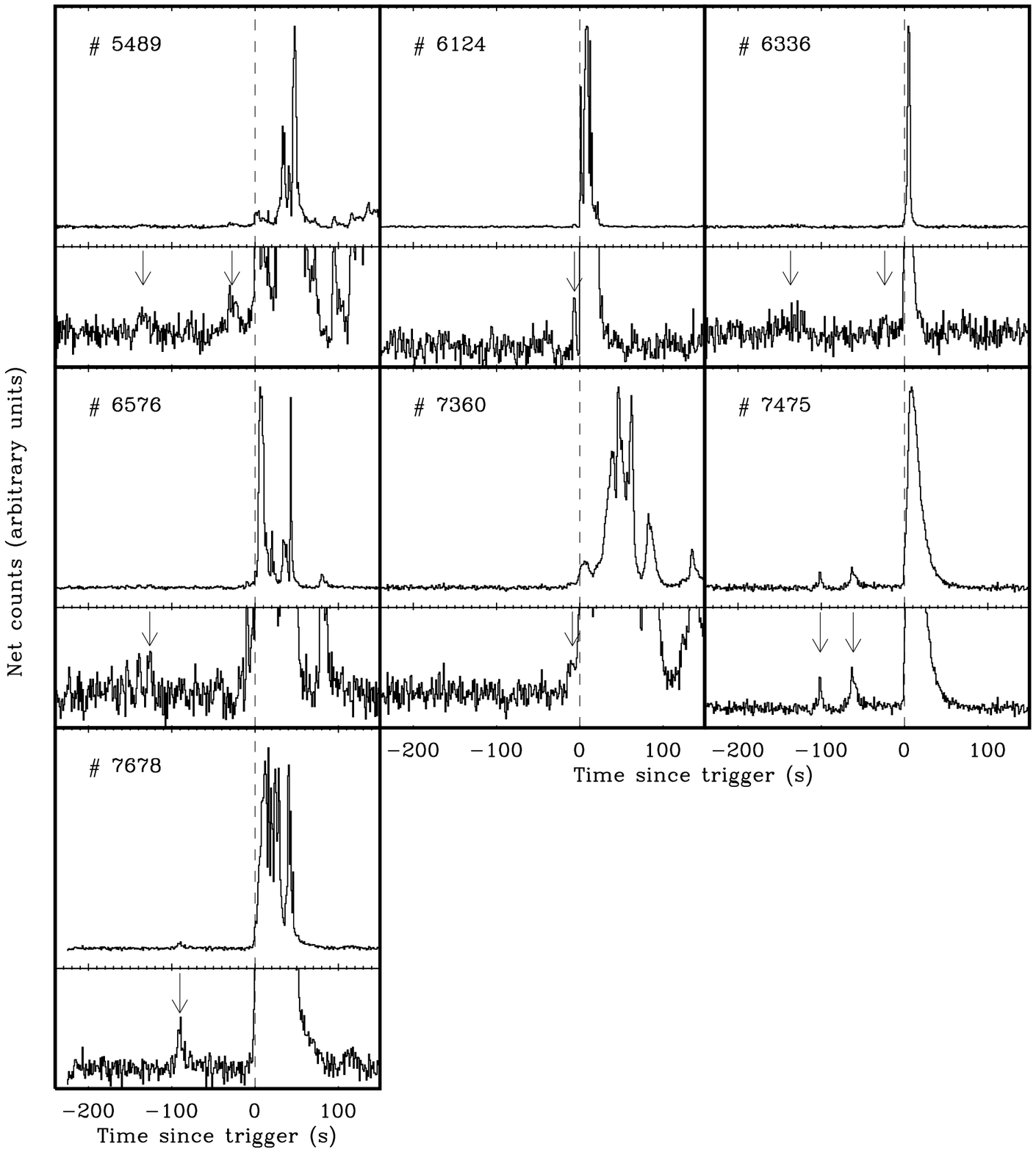,width=\textwidth}
\setcounter{figure}{0}
\caption{{continued}}
\end{figure*}

\subsection{Detector Selection}
\label{sec:bright}

The first step of the analysis is dedicated to the selection of the
detectors more sensitive to photons coming from the direction of the
considered GRB. To this goal, we first added together the 4 channels
of each detectors, then performed a polynomial background fit to the 8
detectors individually and independently, generating the background
subtracted light curve of each detector. We then computed the net
counts of the burst in each detector and selected the detectors in
which the GRB is brighter.

There was no fixed rule on the number of detectors. In some cases a
single detector was dominant, while in other cases the signal was
dominated by four detectors. The subsequent analysis was performed on
the summed data from the selected detectors. The selected number
of detectors is reported in the second column of Tab.~\ref{tab:prec}.

\subsection{Wavelet detection}

In this stage only data in the interval
$t_{\rm{GRB}}-262.144-\epsilon<t<t_{\rm{GRB}}-\epsilon$ were
considered, where $\epsilon$ is a small interval (usually 1 or 2 s)
chosen in order to avoid a steep rise of the count-rate as the GRB
trigger is approached. The number 262.144 is chosen in order to have
256 time bins of 1.024 seconds each (the resolution of the {\tt
DISCLA} data).

The issue is then to search for any emission component on top of a
scarcely predictable variable background. This is a non trivial task,
since the definition itself of emission exceeding a variable
background is non straightforward. More so if we consider that the
BATSE background in the soft channels (where we expect our precursor
to be brighter) is known to suffer from non Poissonian noise
(Connaughton 2002). 

Giblin et al. (1999; see Connaughton 2002 for a full description)
describe a procedure to estimate the background of the BATSE detectors
at any given time. They adopt data from the satellite when it is
positioned at the most similar geo-magnetic coordinates, i.e. 15
orbits ($\sim84000$~s) before and after the time of interest. They
show that the average of the two rates gives a consistent estimate of
the background rate at the time of interest, provided that: i) the
data coverage is continuous in that interval; ii) there is no
serendipitous source in the time intervals adopted for the background
and iii) that the softest channel is not considered, since it is
affected by unpredictable non-Poissonian noise.

\begin{table*}
\caption{{Summary of the detection and control analysis of the samples
considered. Note that all the number of detections are for GRB, i.e. a
burst with a multiple precursor counts only 1. On the other hand,
since some bursts had both a confirmed precursor and a spurious one
(based on the direction of arrival of the photons), the number of
spurious plus confirmed detections is larger than that of total
detections for the GRB sample. The average duration of spurious
precursors (50~s) is longer than that of confirmed ones ($10.7$~s).}
\label{tab:summ}}
\begin{tabular}{l||cc}
                              & GRB sample & Control Sample \\ \hline
Number of lightcurves         & 133        & 207            \\
Number of detections/fraction & 36/27\%    & 22/11\%         \\
Number of spurious det./fraction  & 14/11\%    & 22/11\%         \\
Number of confirmed det./fraction  & 25/19\%    & 0/0\%
\end{tabular}
\end{table*}

We have tested this method and agree with the conclusions of
Connaughton (2002), finding it not suitable for our analysis. On the
one hand, it would lead to a further reduction of the sample due to
the constraints above. On the other hand, it provides a good
subtraction of the slowly evolving background in the soft channels,
but not as accurate for the short time scale fluctuations, leaving
therefore any potential contaminant to our search unaffected. We have
therefore addressed the problem in a different way, adopting a
detection algorithm which automatically corrects for slowly evolving
background. This allows all the contaminants to be included in a first
catalogue, which is then analysed to reject any source whose direction
of arrival does not coincide with that of the GRB. We find {\it a
posteriori} (see below) that this method is quite reliable, leading to
a final version of the precursor catalogue with a low level of
contamination.

\begin{figure*}
\parbox{0.49\textwidth}{\psfig{file=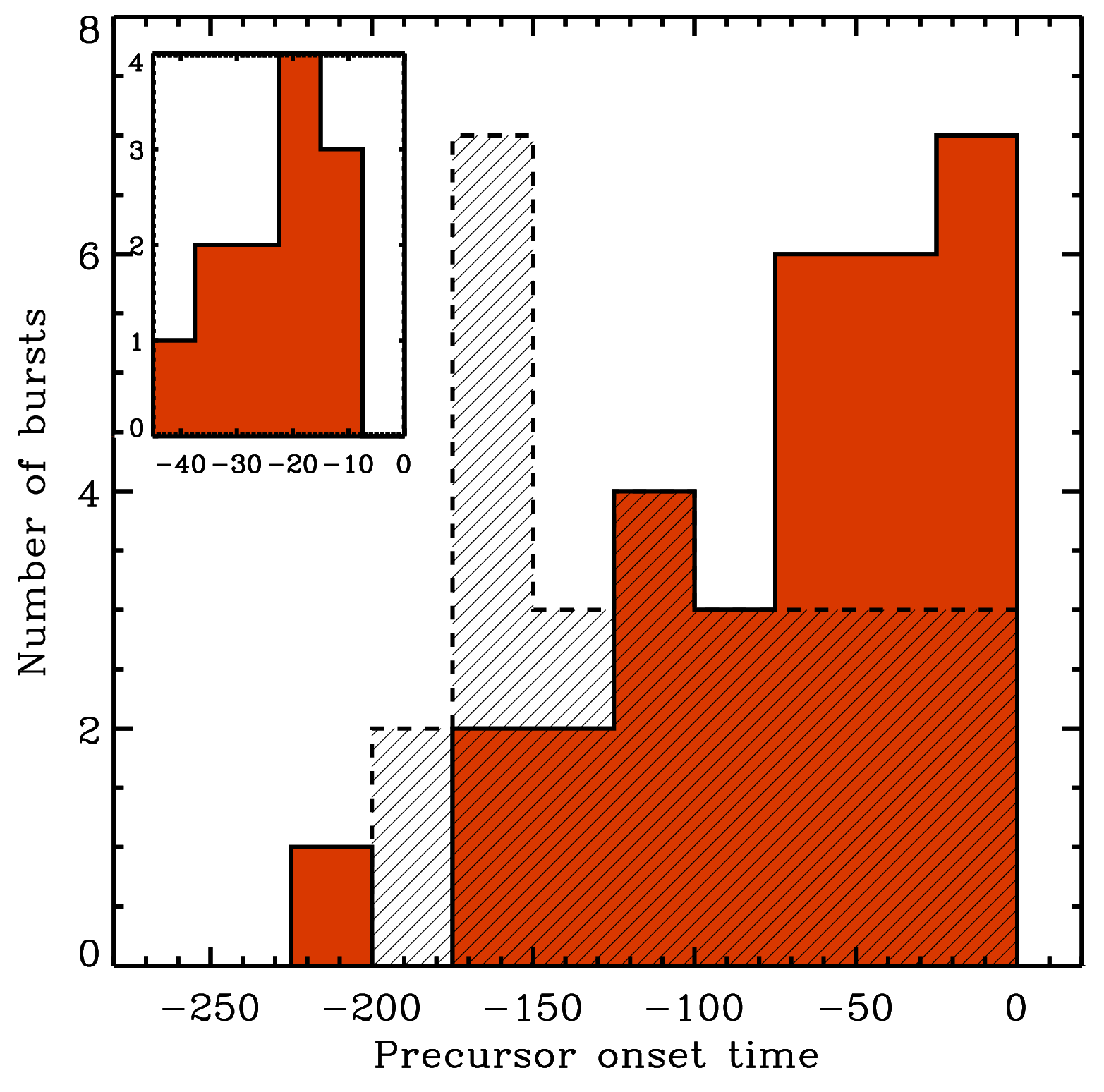,width=\columnwidth}
\caption{{Distribution of the delays between the detected precursors 
and the trigger time of the prompt emission (filled histogram). The
line filled histogram shows instead the distribution of the times at
which spurious emission is detected in the control sample. The latter
distribution is consistent with uniformity, while the true precursors
are more likely to be detected close to the GRB. The inset show a
higher resolution zoom of the distribution at small delays,
emphasising the paucity of precursors with delays $\Delta{}t\le15~s$,
very likely due to an incompleteness of our catalogue.}
\label{fig:delhist}}}
\parbox{0.49\textwidth}{\psfig{file=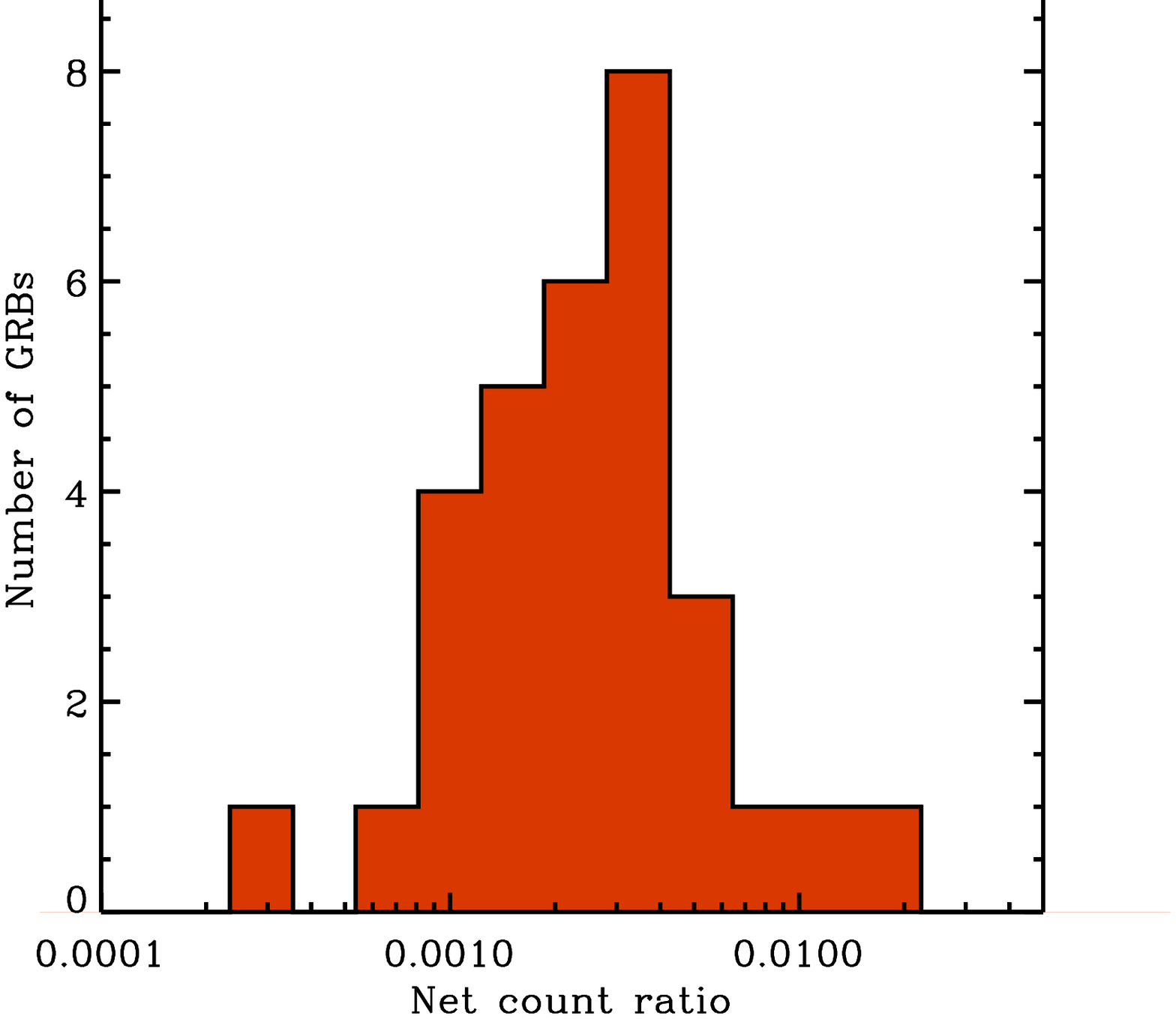,width=\columnwidth}
\caption{{Distribution of the ratio of the precursor net 
counts over the main GRB net counts.\vspace{2.4cm}}
\label{fig:fgfp}}}
\end{figure*}

The search for emission episodes in the selected time interval was
performed with the use of a wavelet transform algorithm. Wavelet
transform have multiple advantages with respect to sliding boxes in
this case. Firstly, they perform an in-situ background subtraction,
being insensitive to large-scale background variations. Secondly, they
perform a multi-scale search, so that excesses are detected
independently from their duration.

We adopted a reduction to one dimension of the two dimensional Mexican
hat wavelet developed by Lazzati et al. (1999) for the analysis of
ROSAT-HRI observations (Campana et al. 1999; Panzera et al. 2004). The
wavelet mother is built as the subtraction of two Gaussian, the
negative one being wider by a factor $\sqrt{2}$. The time series is
transformed and candidate sources are selected as peaks in the wavelet
space at each scale in the time interval 200 seconds before the burst
trigger. The statistical significance of candidate sources is assessed
via simulations. For each burst, we simulate $10^4$ mock Poissonian
time series with the same large scale evolution (modelled with a
polynomial function). We discard any peak in the WT of the real data
which does not exceed the largest peak, in the same scale, of all the
simulations, so that any candidate source has a probability $<10^{-4}$
of being spurious. Finally, we cross correlate the catalogue in order
to get rid of sources detected simultaneously at different scales in
the same position.

At this stage we are left with a catalogue of non-Poissonian excesses,
characterised by a delay time from the GRB trigger and a rough
estimate of their duration (the scale at which the source was detected
at the highest signal to noise ratio). We cannot yet conclude that
these are precursors associated to the forthcoming GRB emission, since
we know that such non-Poissonian events are present in the BATSE
detectors even without any association with a GRB. In order to rid the
catalogue of spurious detections we compute the net counts of any
candidate source in the 8 detectors independently. This was
accomplished by fitting Gaussian functions to the excess. First, on
the best signal-to-noise detectors, the Gaussian is fitted with all
parameters (centroid, width and normalisation) free, together with a
third degree polynomial for the background modelling. Secondly, the
same Gaussian, with fixed centroid and width, is fitted to the
remaining detectors, leaving also the background free to adjust
independently in each case. We then compare the relative counts of the
candidate precursor in the different detectors with those of the
prompt GRB emission. If the relative rations between the detectors
(including upper limits) are in agreement, this means the candidate
precursor emission comes from a direction in agreement with the
direction from which the main GRB was detected. In statistical
terms, in agreement means that the reduced $\chi_\nu^2\le2$ and that
no $3\sigma$ upper limit is violated.  In this case the candidate
precursor is elevated to the rank of confirmed precursor. Otherwise it
is discarded. As it is summarised in Tab.~\ref{tab:summ}, this
procedure led us to identify 25 GRBs with precursors out of a
contaminated catalogue of 36 GRBs with excesses. In 3 cases, the GRB
showed both accepted and rejected episodes of emission. A word of
caution should be spent for the possible biases of this
technique. Since weak precursors have larger error bars, it's easier
for them to be in agreement with the prompt emission relative counts
and we expect therefore the contamination to be larger at small
fluxes. This is however a problem of all catalogues: the closer is a
source to the detection limit, the higher is the probability for it to
be spurious. In addition, even if the test is performed on all the 8
detector, a GRB bright in only one detector will have a higher
probability of a spourious precursor.

We still expect this catalogue to be contaminated to some level, since
some precursors are detected only in few detectors (several even in a
single one), so that the comparison with the count ratios of the
prompt is only indicative in these cases. To assess the level of
contamination, we analysed a control sample of lightcurves, not
associated to any known GRB emission, extracted from the {\tt DISCLA}
data 1 day before and 1 day after the analysed GRBs. We ended up with
a control sample of 207 lightcurves, since all the cases in which the
sampling of the interval was not continuous had to be discarded. These
lightcurves were analysed in the same way as before, except for some
inevitable differences. First, the number of detector where the
analysis is performed, was randomly selected between 1 and 4, to mimic
the first step of the procedure. Secondly, the detector ratios were
not computed, since there is no ratio of the prompt emission to
compare them to, and all the detected excesses are in this case known
to be spurious.

\begin{figure*}
\parbox{0.49\textwidth}{\psfig{file=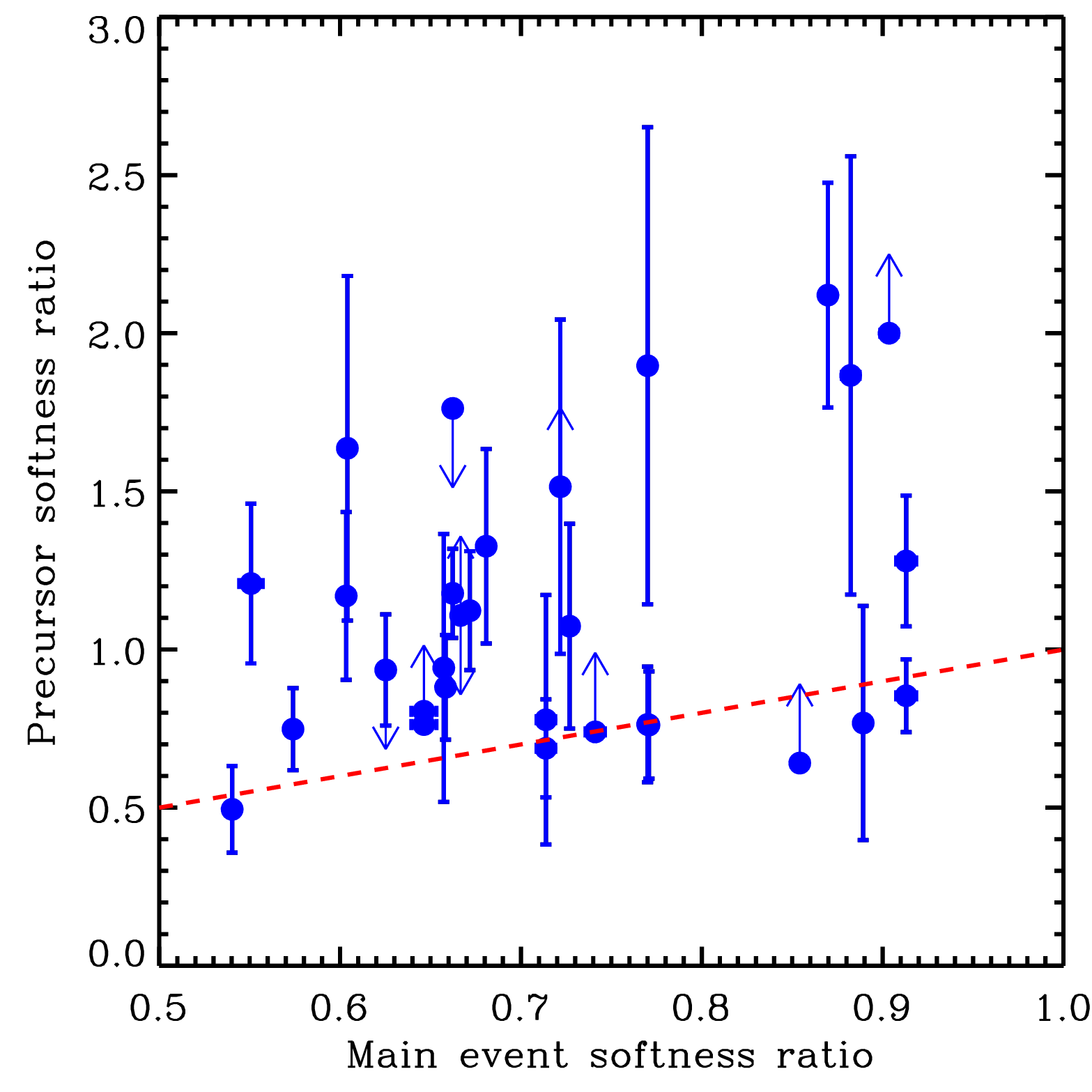,width=\columnwidth}
\caption{{Softness ratio of the precursors vs. softness ratio of the main 
event. The softness ratio is defined as the ratio between the net
counts in the first channel over the net counts in the second
channel. The dashed line shows the locus of points where the two
softness ratios are equal. All the precursors are consistent with
being softer than the GRB to which they are associated.}
\label{fig:sr}}}
\parbox{0.49\textwidth}{\psfig{file=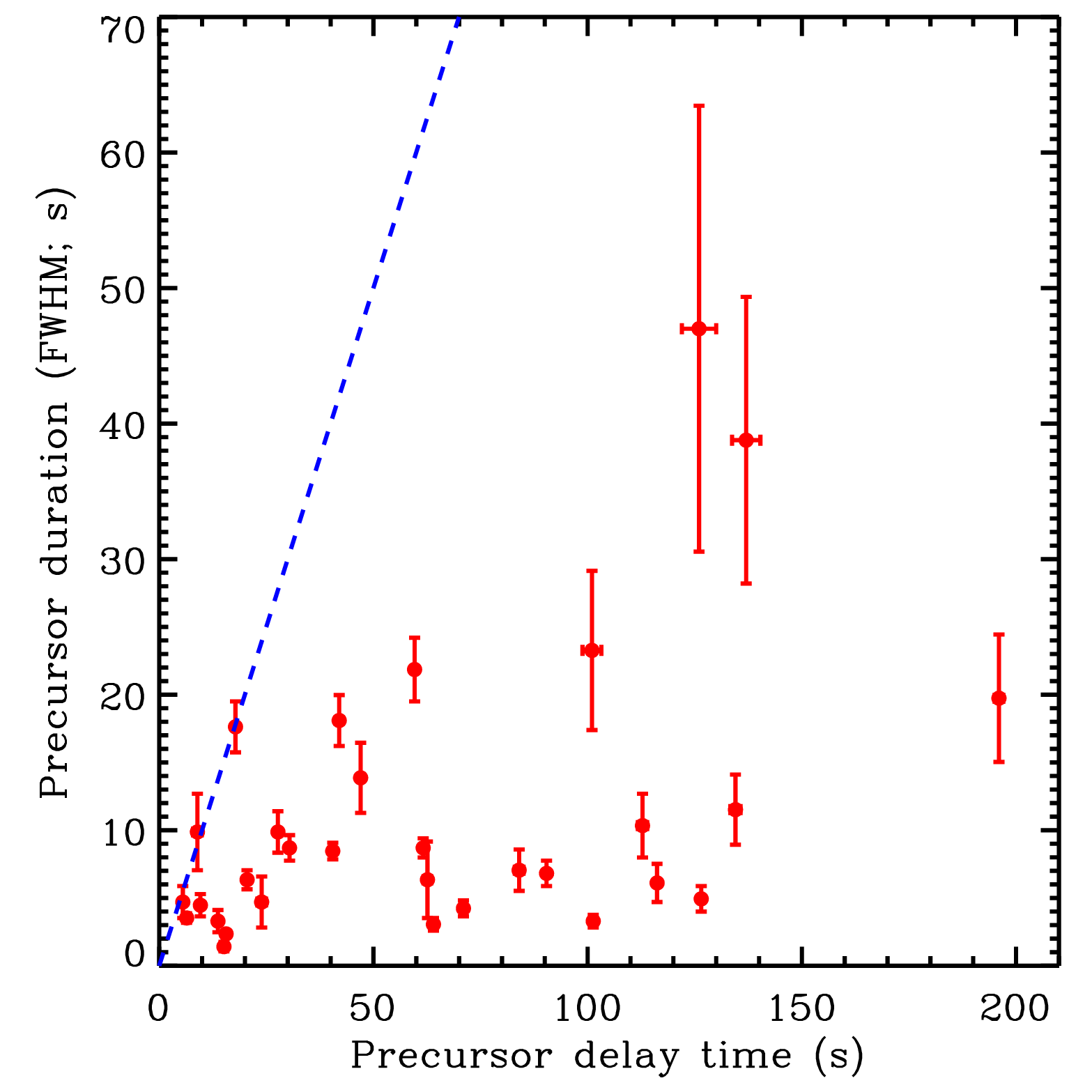,width=\columnwidth}
\caption{{Duration of the precursors (FWHM) vs. delay time. Due to 
the adopted definition of precursor, all the points lay in the region
where the delay is larger than the duration (the dashed line shows the
duration equal delay condition).}
\label{fig:dtDt}}\vspace{0.7cm}}
\end{figure*}

We find that spurious non-Poissonian soft excesses are expected in
$11\%$ of the lightcurves, amounting to 14.6 expected excesses in our
GRB sample. Reassuringly, this is exactly the number of excesses that
were rejected in our analysis, confirming the goodness of our
technique.

The resulting catalogue of precursors is detailed in
Tab.~\ref{tab:prec}. Additional tests on the reality of the 25
selected precursors are discussed below.

\begin{table*}
\caption{{The catalogue of detected precursors. Each precursor is 
labelled with the BATSE trigger number of the main event to which is
associated plus a letter $A$ for the most significant event and $B$ in
case there is a second less significant precursor component.}
\label{tab:prec}}
\begin{tabular}{lcccccccc}
ID    & \# of det. & Delay       & FWHM  & Fluence      & Ratio$^{(a)}$ & $\chi_\nu^2(PL)$ & $\chi_\nu^2(BB)$ & Spectrum$^{(b)}$ \\
      & &(s)         & (s) & ($10^{-8}$~erg~cm$^{-2}$) & ($\times10^{-3}$) & \\ \hline
451A  & 4 & $15.1\pm0.2$  & $1.4\pm0.35$ & $0.32\pm0.1$   & $1.53\pm0.5$ & 0.65 & 3.95 & PL \\
676A  & 2 & $101\pm2.2$   & $23.3\pm5.8$ & $1.2\pm0.3$   &  $2.8\pm0.7$ & - &- &-  \\
1141A & 1 & $113.\pm1.1$  & $10.3\pm2.3$ & $1.3\pm0.36$  & $2.2\pm0.6$ & - &- &- \\
1625A & 2 & $30.4\pm0.4$  & $8.7\pm1$   & $4.3\pm0.6$   & $3.7\pm0.5$ & 2.83 & 3.22 & NO \\
2067A & 3 & $42\pm0.8$   & $18\pm2$    & $4.0\pm0.5$   & $4.5\pm0.6$ & 8.98 & 0.16 & BB \\
2110A & 3 & $71\pm.5$    & $4.2\pm0.6$  & $1.0\pm0.2$  & $2.5\pm0.6$  & - &- &-  \\ 
2228A & 2 & $126\pm4$    & $47\pm16$   & $0.9\pm0.4$  &  $3.1\pm1.5$ & - &- &-  \\
2316A & 3 & $47\pm0.5$   & $14\pm2.6$  & $3.1\pm0.8$   & $5.8\pm1.5$ & 0.31 & 11.32 & PL \\
2371A & 2 & $84\pm1$    & $7.1\pm1.5$  & $0.4\pm0.1$  & $2.3\pm0.7$ & -& - & - \\
2798A & 3 & $40.5\pm0.3$ & $8.5\pm0.6$  & $4.6\pm0.45$ & $2\pm0.2$ & 25.5 & 0.37 & BB \\
2798B & 3 & $9.6\pm0.5$  & $4.5\pm0.8$  & $0.76\pm0.26$ & $0.33\pm0.11$ & - &- &- \\
3115A & 1 & $5.5\pm0.4$  & $4.7\pm1.2$  & $1.0\pm0.2$  & $3.8\pm0.8$  & - &- &-  \\
3241A & 3 & $15.6\pm.2$  & $2.3\pm0.2$  & $1.2\pm0.2$  & $2.4\pm0.3$  & - &- &-  \\
3245A & 2 & $59.6\pm0.7$ & $22\pm2$    & $9.2\pm1.2$  & $3.5\pm0.5$  & 0.29 & 21.4 & PL \\
3345A & 3 & $20.5\pm0.3$ & $6.3\pm0.7$  & $3.0\pm0.4$ & $9.3\pm1.4$   & 0.45 & 11.24 & PL \\
3345B & 3 & $116\pm1$   & $6.1\pm1.4$  & $1.3\pm0.4$  & $4.0\pm1.2$  & 0.01 & 2.28 & PL \\
3408A & 2 & $64\pm0.2$  & $3.0\pm0.5$  & $1.2\pm0.2$  & $1.1\pm0.2$  & 1.6 & 4.34 & PL \\
3408B & 2 & $13.7\pm0.3$ & $3.3\pm0.8$ & $0.6\pm0.25$ & $0.6\pm0.25$ & 1.3 & 11.73 & PL \\
3489A & 4 & $17.8\pm0.6$ & $17.7\pm1.9$ & $8\pm1$    & $16\pm2$    & 2.85 & 10.64 & NO \\
3523A & 2 & $196\pm1$   & $20\pm5$    & $13.\pm3$   & $3.3\pm0.7$  & 0.33 & 17.02 & PL \\
3765A & 2 & $62.6\pm0.9$ & $6.4\pm2.8$ & $0.5\pm0.2$  & $1.0\pm0.4$ & 0.05 & 2.77 & PL \\
5489A & 2 & $27.7\pm0.7$ & $10\pm1.5$  & $1.0\pm0.2$  & $2.0\pm0.4$ & - & - & - \\
5489B & 2 & $134.5\pm1.2$ & $11.5\pm2.6$ & $0.7\pm0.2$ & $1.3\pm0.4$ & - &- &- \\
6124A & 4 & $6.4\pm0.15$ & $3.5\pm0.4$  & $4.7\pm0.6$  & $4\pm0.5$ & 1.76 & 10.21 & PL \\
6336A & 2 & $137\pm3$   & $39\pm10$   & $6.0\pm2.5$  & $10\pm4$ & - & - & - \\
6336B & 2 & $24\pm1$    & $4.7\pm1.9$ & $0.8\pm0.4$  & $1.4\pm0.7$ & - & - & - \\
6576A & 3 & $126.5\pm0.4$ & $5\pm1$   & $0.8\pm0.2$  & $0.8\pm0.2$ & 0.96 & 11.95 & PL \\
7360A & 2 & $9\pm1$     & $10\pm3$   & $0.7\pm0.2$  & $1.2\pm0.4$ & 0.72 & 1.05 & PL/BB \\
7475A & 1 & $61.6\pm0.25$ & $8.7\pm0.7$ & $2.6\pm0.3$ & $4.9\pm0.5$ & 1.73 & 21.72 & PL \\
7475B & 1 & $101.3\pm0.15$ & $3.3\pm0.5$ & $0.9\pm0.2$ & $1.7\pm0.3$ & 1.19 & 14.23 & PL \\
7678A & 4 & $90.4\pm0.4$ & $6.8\pm0.9$  & $1.9\pm0.4$ & $1.6\pm0.3$ & 1.42 & 9.83 & PL
\end{tabular}

\noindent
$^{(a)}$ Ratio between the main GRB emission and the precursor net
counts in the whole BATSE sensitivity band.

\noindent
$^{(b)}$ Best fit spectral shape. PL=power law; BB=black body; NO=no
model yielded an acceptable $\chi^2$; - = not enough points for a
meaningful fit.
\end{table*}

\subsection{Characterisation}

The characterisation of the detected precursors is performed by
fitting Gaussian profiles to the data. Given the low signal-to-noise
ratio of the sources, it is not possible to study their structure in
further detail, and a symmetric fitting function like a Gaussian is
always adequate. The width and centroid of the precursor is determined
by fitting the highest signal-to-noise data, i.e. the sum of the first
three channels of the brightest detectors
(Sect.~\ref{sec:bright}). The fit are always performed on the data
without background subtraction. The background component is described
by adding a 3$^{\rm{rd}}$ degree polynomial to the Gaussian.

Extracting a spectrum from the BATSE LAD data for such faint sources
is hopeless. We have therefore extracted a very-low resolution
spectrum of the time-integrated precursor by re-normalising the four
channel fluences of the main events, derived from the BATSE
catalog, to the net-counts of the precursor.  In this way we obtain a
three point spectrum for most of the spectra. In the high energy
channel the count-rate is always to low to yield any significant
detection, so that only a upper limit can be obtained. In some cases
also at softer energies the precursor is too weak to be detected. In
these cases, when only two or less spectral points are available, we
did not attempt to characterise the spectrum since any model can fit.
Given the roughness with which the three channel spectra are derived,
we attempted only to fit power-laws and black body spectra. The model
spectra were integrated in the broad energy bin before being fitted to
the data. The statistical significance of the fit and the best
model were evaluated by means of $\chi^2$ statistics, see
Tab.~\ref{tab:prec}. All the uncertainties quoted in this paper are
at the $1$-$\sigma$ level.

\section{Results}

Table~\ref{tab:prec} reports the results of our search. We find
precursor activity, in the 200 seconds preceding the GRB trigger, in
25 burst out of 133 searched, for a total of 31 individual precursors
events. Lightcurves of the bursts in which precursor activity has been
detected are shown in Fig.~\ref{fig:atlas}. We conclude therefore that
at least about $20\%$ of the bright BATSE GRBs are characterised by a
precursor activity to some extent. We say ``at least'' since, given
the problems in the definition of a precursor (see Sect.~1), it is
unavoidable that our search procedure misses some cases. As examples,
consider a very bright precursor or a very small delay precursor (as
predicted by theory). The former will trigger as a burst and will be
excluded from our search, while the second will be mixed up with the
main event, and therefore undetectable. This in addition to precursor
events too weak to be detected.

\begin{figure*}
\parbox{0.49\textwidth}{\psfig{file=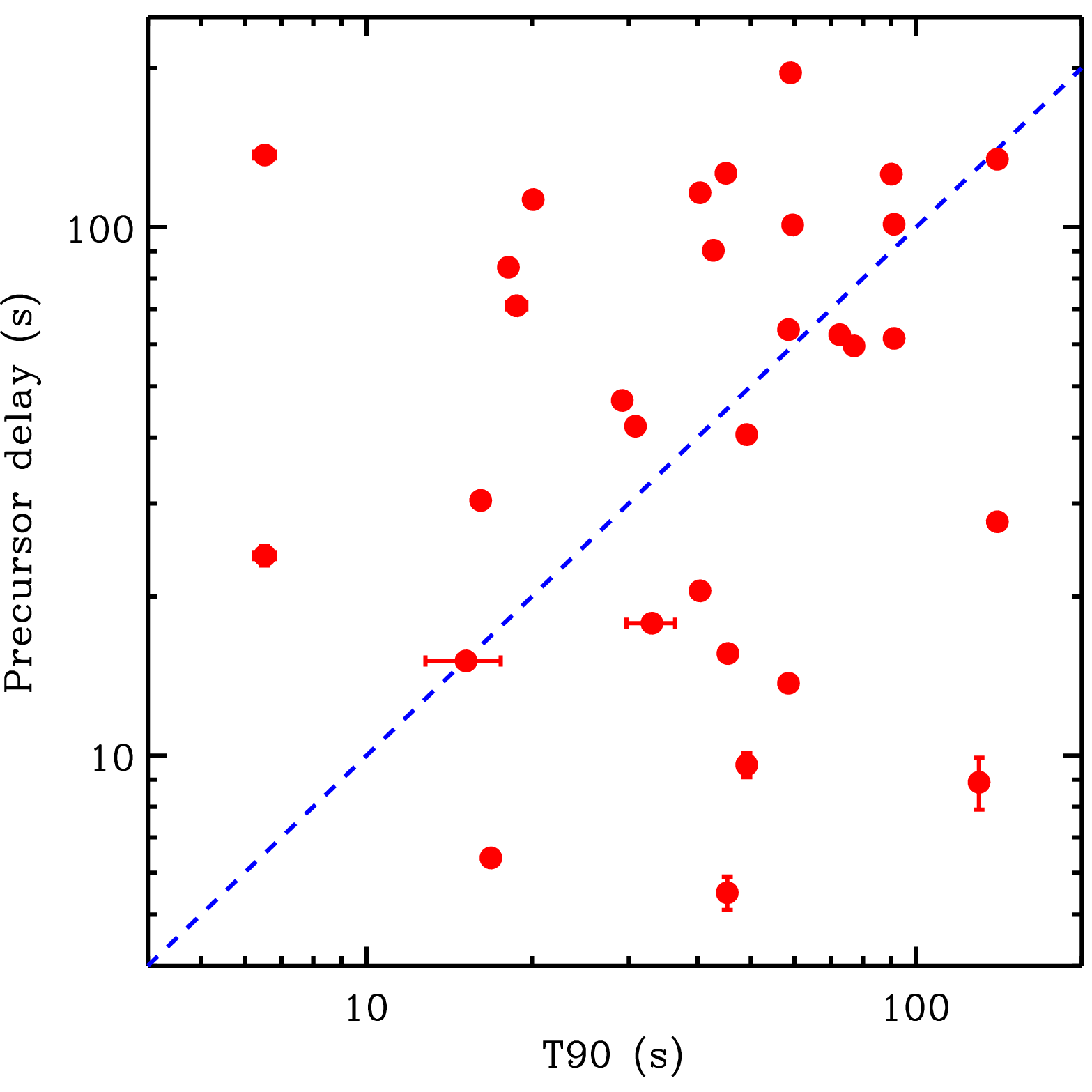,width=\columnwidth}
\caption{{Precursor delay vs. duration of the main GRB emission.  No
correlation is apparent in the data. The correlation coefficient
is $r=0.082$.}
\label{fig:Dtt90}}}
\parbox{0.49\textwidth}{\psfig{file=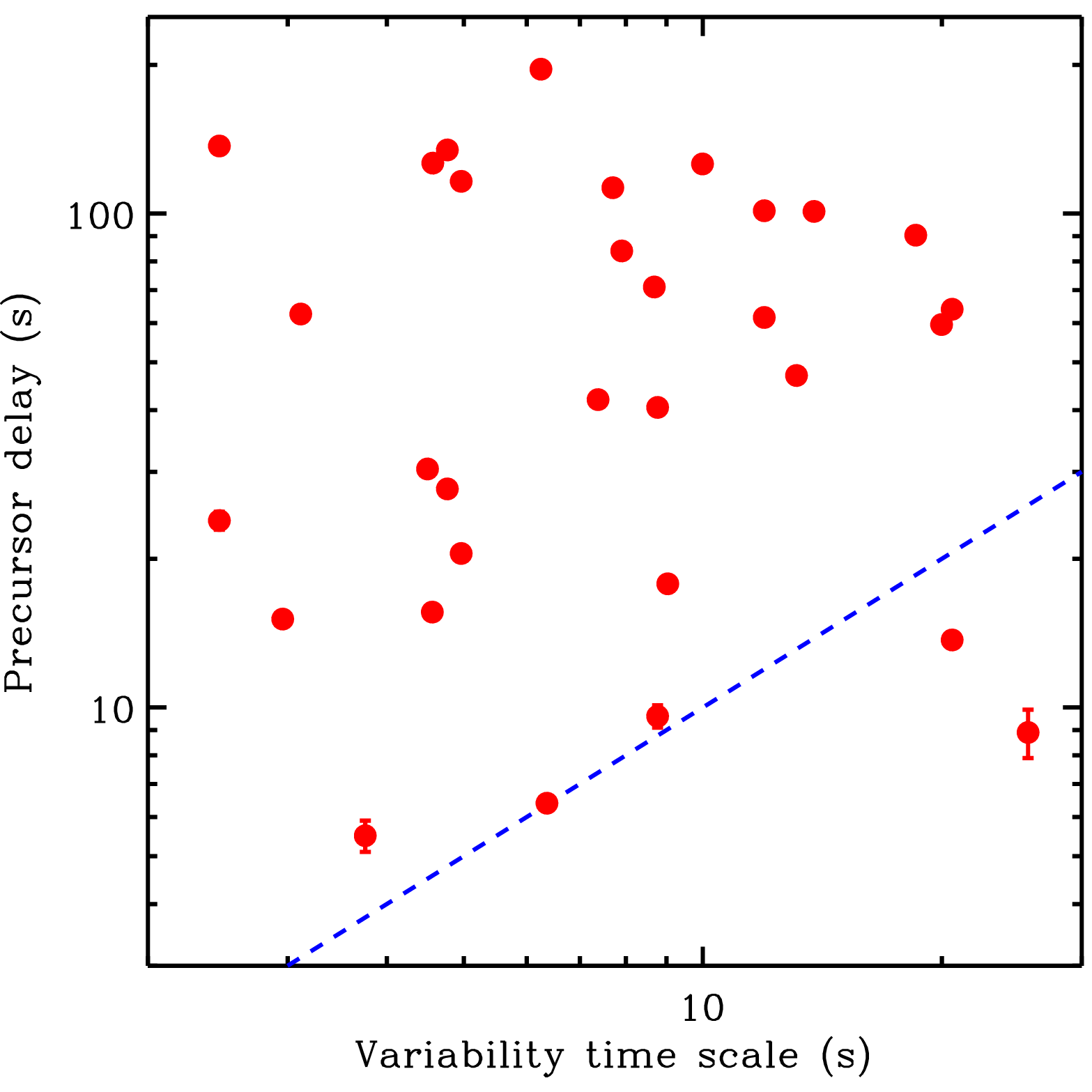,width=\columnwidth}
\caption{{Precursor delay vs. variability time scale of the main GRB
emission. No correlation is apparent in the data.The correlation
coefficient is $r=0.033$.}
\label{fig:Dttv}}}
\end{figure*}

The distribution of delay times of the detected precursors is shown
with the filled histogram in Fig.~\ref{fig:delhist}. The onset time of
the precursor is here defined as the centroid of the best fit Gaussian
minus twice its $\sigma$. For comparison, with a line filled
histogram, we show the distribution of the spurious precursors
detected in the control sample. As expected, spurious detections are
uniformly distributed in the whole interval, while the ``real''
precursors are more likely to be detected close to the trigger time of
the GRB. The inset shows a higher resolution distribution for the
small delay region, to underline the paucity of precursors with small
time delay. As discussed above, this is likely to be due to an
incompleteness of our catalogue rather than to a real paucity of
cases. Nonetheless, it is surprising to note that the average delay
time is of the order of several tens of seconds, while theory predicts
delays of the order of few seconds at most, if not of a fraction of a
second (see below).

Unfortunately no redshift information is known for the GRB sample
where we perform the search and we cannot compute the energy that is
contained in the precursors. We characterise therefore the energetics
of the precursors through the ratio of their net counts to that of the
main GRB to which they are associated. We should however keep in mind
that the beaming of GRB and precursor photons may well be
different. The distribution of this ratio is shown in
Fig.~\ref{fig:fgfp}. The precursors we detect contain on average a
sizable fraction of a per cent of the total counts of the event. These
precursors are therefore quite energetics especially if, as predicted
in some models, their beaming angle is wider than that of the main
emission.

It may be argued that what we are detecting is not precursor activity
but merely the beginning of the prompt emission. If we compare the
count ratios in different channels this seems not to be the case. In
Fig~\ref{fig:sr} we compare the softness ratio of the precursors with
that of the integrated prompt emission. The softness ratio is here
defined as the ratio of the first BATSE channel net counts over the
second channel net counts. We find that all the precursors are softer
than the time-integrated prompt emission. A similar softness (the
dashed line in Fig.~\ref{fig:sr}) has a vanishing probability of
$P=3\times10^{-6}$, and can therefore be rejected at about $5\sigma$.
This is in striking contradiction with what is expected from the
hard-to-soft evolution usually found in GRB spectra (Ford et al. 1995;
Frontera et al. 2000). Note again that we do not compare the precursor
to the beginning of the burst, which is known to be particularly hard,
but to the average softness. Is seems therefore that the emission we
single out before the burst is indeed something with a different
origin.

In Fig.~\ref{fig:dtDt} we plot the duration (the full width at half
maximum FWHM) of each precursor versus its delay time. Since we
require a decrease in flux before the trigger in the precursor
definition, all the points lay in the region of the plot where the
precursor FWHM is less than its delay. The dashed line shows the locus
of points where the equality is realised. The comparison of this
figure with Fig.~\ref{fig:delhist} allows us to roughly estimate the
number of lost precursors due to the above condition. Under the
assumption that half of the precursors with duration $<20$~s are lost
(Fig.~\ref{fig:dtDt}), we conclude that we miss $\sim9$ precursors in
our search (Fig.~\ref{fig:delhist}), increasing the fraction of GRBs
with precursors to $\sim25\%$. Of course, this estimate is based
on the assumption, difficult to test in a robust way, that the lack of
precursors with similar delay and duration longer than 20 seconds is a
real effect and not a detection bias. This estimate should therefore
be taken as tentative. To these, we shall add those missed due to
their short delay (inset in Fig.~\ref{fig:delhist}), more difficult to
quantify.

We now want to compare the properties of the precursors with those of
the main event to see is any correlation exists. First, we compare in
Fig.~\ref{fig:Dtt90} the delay of the precursors with the burst
duration (its $T_{90}$). In Fig.~\ref{fig:Dttv}, instead, we compare
the precursor delay with a measure of the burst variability time
scale. We measure it as the half width of the auto-correlation
function (following Borgonovo 2004). In both figures, there appear to
be no correlation whatsoever. This is confirmed by running statistical
tests on the data.

\begin{figure*}
\parbox{0.49\textwidth}{\psfig{file=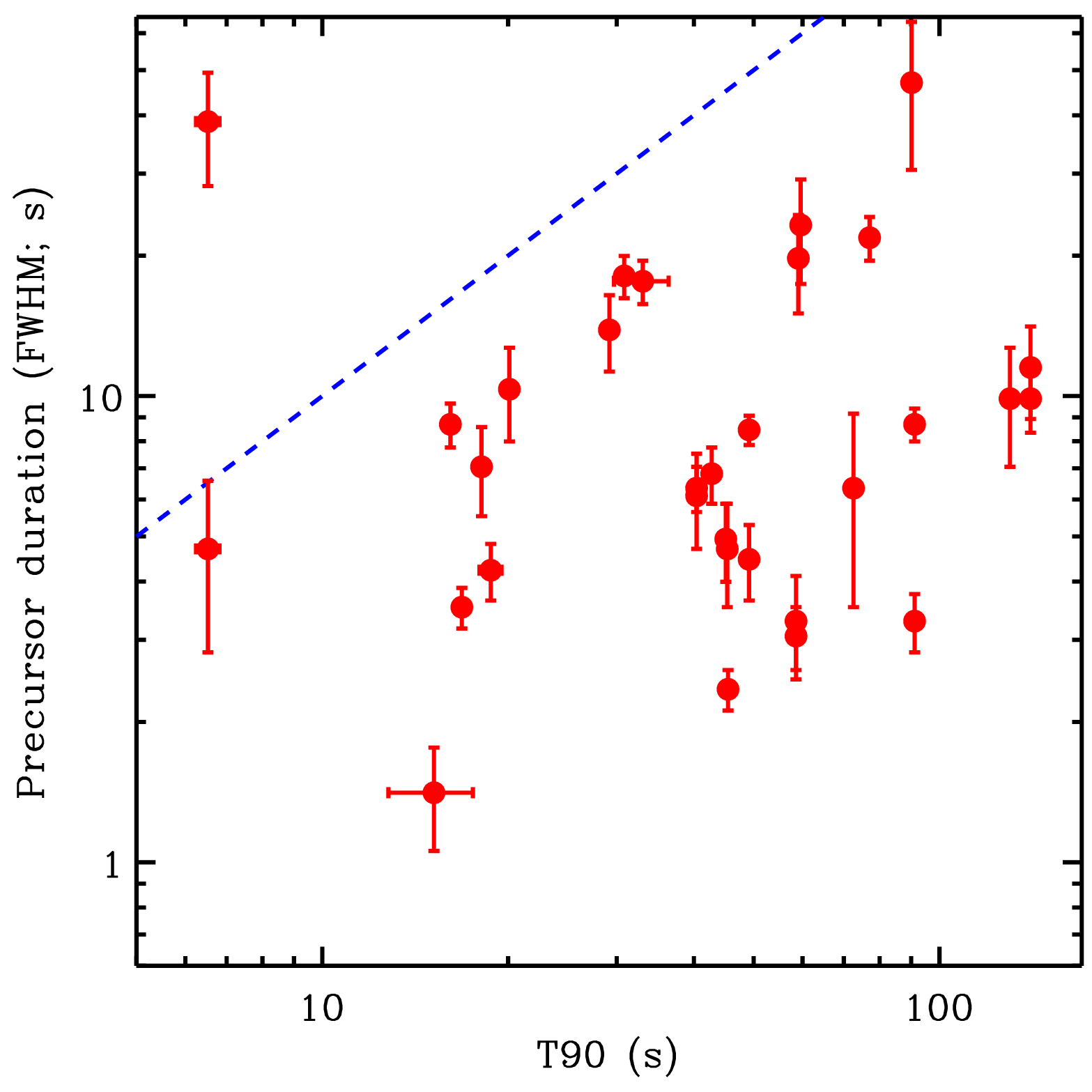,width=\columnwidth}
\caption{{Precursor duration vs. duration of the main GRB emission. A
mild correlation may be present, but it is not statistically
compelling.  The correlation coefficient is $r=0.11$.}
\label{fig:dtt90}}}
\parbox{0.49\textwidth}{\psfig{file=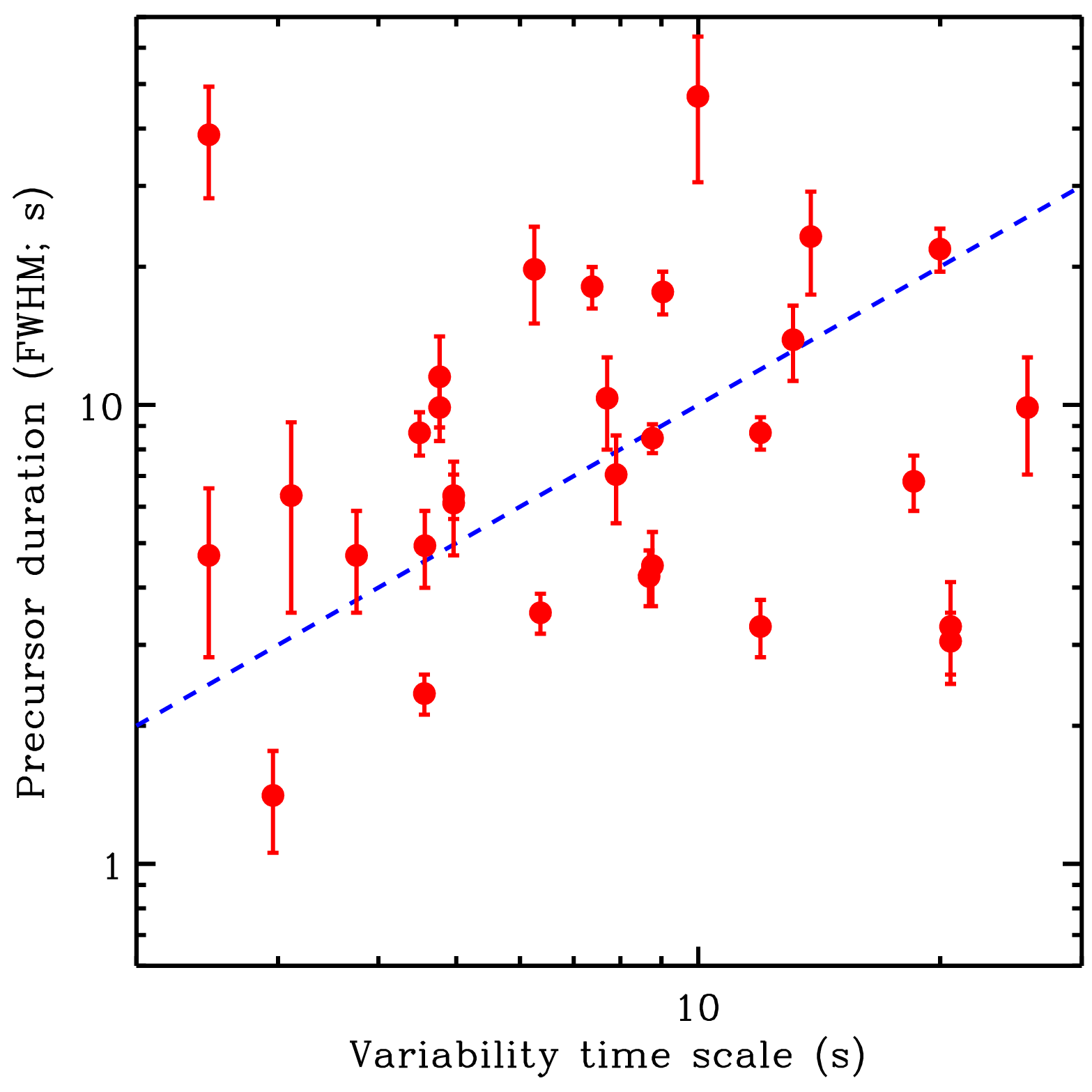,width=\columnwidth}
\caption{{Precursor duration vs variability time scale of the main GRB
emission. A mild correlation may be present, but it is not
statistically compelling.  The correlation coefficient is $r=0.13$.}
\label{fig:dttv}}}
\end{figure*}

In Figs.~\ref{fig:dtt90} and~\ref{fig:dttv} we perform the same
comparison but using the precursor duration (its FWHM) instead of its
delay. Even though the small number statistics does not allow us to
draw and definitive conclusion, it appears that the precursor duration
is more closely correlated to the temporal properties of the main
burst emission. In particular, there is only one case in which the
precursor lasts more than the main emission.

\section{Discussion}

We have analysed a sample of bright long BATSE GRB lightcurves to
search for weak emission episodes taking place before the burst
trigger or precursors. We define them as any count variation localised
in time taking place before the burst trigger and coming from a
location in the sky consistent with the direction of the main GRB
event. This latter condition is verified through the consistency of
the relative brightnesses of the burst and precursor in the eight
BATSE detectors. We find that at least $\sim20\%$ of the analysed
bursts do have weak precursors. The precursor are weak, containing
only a fraction of a per cent of the total counts of the event. Their
properties do not correlate with the main event properties (as found
also by Koshut et al. 1995; albeit under a largely different
definition of precursor activity). We only find a mild correlation of
the precursor duration (not its delay) with the burst $T_{90}$ and its
variability time scale.

Two important properties of the precursors are surprising. First their
delay is long. Typical delays are of tens of seconds (see
Fig.~\ref{fig:delhist}), but precursors with up to $\sim200$ seconds
delays are observed. If the precursor is associated in some way with
the fireball initial release from the central compact object, and the
$\gamma$-rays of the main GRB are released at a radius $R_\gamma$, at
which the fireball has a Lorentz factor $\Gamma_\gamma$, we expect a delay:
\begin{equation}
\Delta t = \frac{R_\gamma}{2c\,\Gamma_\gamma^2} = 0.02\,R_{\{\gamma; 13\}}
\,\Gamma_{\{\gamma; 2\}}^{-1} \;\;\; {\rm s}
\end{equation}
between the precursor onset and the trigger\footnote{Here and in the
following we define any quantity $Q=10^x\,Q_x$.}. For standard
internal shock parameters, the delay time is of the order of a
fraction of a second. One may be tempted to vary the fiducial
parameters in order to have a longer delay. This seems not to be the
right way. As a matter of fact, a minimum variability time scale is
associated to any radius and Lorentz factor by the curvature time scale
(Fenimore et al. 1996):
\begin{equation}
t_{\rm{var}} \ge t_{\rm{curv}}=\frac{R_\gamma}{2c\,\Gamma_\gamma^2} =
\Delta t
\end{equation}
where $t_{\rm{var}}$ is the observed variability time scale.  If the
above interpretation for the delay were correct, we would therefore
expect a relation $t_{\rm{var}}\ge\Delta{}t$, in striking
contradiction with the results of Fig.~\ref{fig:Dttv}, which seems to
underline an opposite relation.

If we consider progenitor precursors, i.e. precursors due to the
interaction of the jet with its progenitor, we arrive to fairly
similar conclusions. This is due to the fact that, again, the
precursor is generated with the jet. There seem to be only two
possible way out to this apparent contradiction. On the one hand, the
curvature time scale may not apply. On the other, the precursor may be
generated before the jet. In the first case, one would need to assume
a fragmented fireball (Heinz \& Begelman 1999; Dar \& De Rujula 2000;
2004), or that the emission comes from localised spots (Lyutikov \&
Blandford 2004) in an otherwise continuous fireball.  Also, an
external shock onto a fragmented interstellar-medium may solve the
variability problem (Dermer, B\"ottcher \& Chiang
1999). 

Alternatively, one may associate the precursor with something taking
place {\emph before} the jet is launched. In a binary merger scenario
this may be associated to the first interaction of the binary
system. In the hypernova scenario, it is harder to find any source of
high energy emission before the jet is released, especially if we
consider that the precursors have non-thermal spectra and must
therefore be generated in an optically thin environment. A final
possibility is that the first part of the jet, for some reasons that
presently escape our understanding, is not radiative and not produce
GRB emission.

A second surprise, as anticipated, comes from the spectrum of the
precursors. All the precursor activity that has been predicted in the
various models is characterised by thermal spectra. In our sample of
19 precursors for which the data quality allowed a spectral
characterisation, only two precursors are characterised by thermal
emission, plus one with a dubious classification. In fact this is not
a completely new issue, since two of the most well studied precursors,
those of GRB~011121 (\bsax; Piro et al. in preparation) and of
GRB~030329 (HETE2; Vanderspek et al. 2004) are also characterised by
non-thermal emission.

\section{Summary and Conclusions}

We have shown that a sizable fraction of bright GRBs are characterised
by weak but significant precursor activity. These precursors have a
delay time from the main GRB which is surprisingly long, especially if
compared to the variability time scale of the burst itself. The
precursor emission, contrary to model predictions, is characterised by
a non-thermal spectrum, which indicates that relativistic electrons
are present in the precursor emission region and that this region is
optically thin. Unfortunately no redshift information is available for
the GRB sample considered, so that it is not possible to estimate the
energy involved in the precursor activity.

Future missions, such as Swift, will provide a large sample of GRB
lightcurves with redshift where a similar precursor search can be
performed. In addition, imaging capabilities will allow a more
effective identification of activity related to the burst itself. This
will allow us to gain a deeper insight on these precursors and to
exploit all their power in the understanding of the GRB phenomenon.

\section*{Acknowledgements}
I thank Giancarlo Ghirlanda for useful discussions on the use of BATSE
data and the anonymous referee for his/her useful comments. This
research has made use of data obtained from the High Energy
Astrophysics Science Archive Research Center (HEASARC), provided by
NASA's Goddard Space Flight Center. This work was financially
supported by the PPARC postdoctoral fellowship PPA/P/S/2001/00268.

\end{document}